\documentclass[12pt,preprint]{aastex}
 
\begin{document}
 
\title{Toroidal Atmospheres around Extrasolar Planets}
 
\author{R. E. Johnson}
\affil{ Engineering Physics and Astronomy Departments, University of Virginia,
Charlottesville, VA 22904}
\email{rej@virginia.edu}
 
\author{P. J. Huggins}
\affil{ Physics Department, New York University, 4 Washington Place,
 New York, NY 10012}
\email{patrick.huggins@nyu.edu}

\begin{abstract}
Jupiter and Saturn have extended, nearly toroidal atmospheres composed
of material ejected from their moons or rings. Here we suggest that
similar atmospheres must exist around giant extrasolar planets and
might be observable in a transit of the parent star. Observation of
such an atmosphere would be a marker for the presence of orbiting
debris in the form of rings or moons that might otherwise be too small
to be detected.
\end{abstract}
                                                                               
 
\section{Introduction}
 
In addition to their usual spherical atmospheres, giant planets have
highly extended, often toroidal atmospheres that are the gaseous
envelopes of material orbiting the planet as rings or satellites. At
Jupiter and Saturn these atmospheres are produced by the ejection of
atoms and molecules from their moons and rings by out-gassing,
volcanism, sputtering, and meteoroid bombardment. Such 
atmospheres are also likely to be present on giant, extrasolar planets.
 
The sodium cloud at Jupiter, associated with its moon Io, was the
first observation of such an extended atmosphere
\citep{bro74}. Because of its considerable tidal stressing, Io is
volcanically active \citep[e.g.,][]{bag04} producing an atmosphere
that is stripped at a rate of about $10^6$~gr~s$^{-1}$ by the plasma
trapped in Jupiter's magnetosphere \citep{tho04}, a process often
called atmospheric sputtering \citep{joh90,joh94}.  The ejected
neutrals are eventually ionized adding to the trapped plasma in a
self-limiting feedback process \citep{joh93,mcg04}.  The neutral cloud
and its ionized component are confined to a region around Io's orbit
but have a radial scale comparable to the size of Jupiter
(Fig.~1a). Observations using the Hubble Space Telescope have shown
that Saturn has a similar, extended, roughly toroidal atmosphere,
composed primarily of water products \citep{she92,jur02}. In this case
it is the gaseous envelope of a disk of icy debris (Fig.~1b)
apparently produced by venting from the icy moon Enceladus
\citep{wai06,tok05}. This atmosphere is also sustained by sputtering 
of the ice grains in Saturn's tenuous E-ring \citep{jur02} and has an 
extended component formed by charge exchange \citep{joh05,joh06a}. Saturn's 
main rings also have a toroidal atmosphere of molecular oxygen produced 
by the decomposition of ice by the solar UV \citep{joh06b}. Associated 
with these extended neutral atmospheres are magnetically confined 
plasmas that are representative of the source material (Fig.~1c).
 
The toroidal atmospheres observed around Jupiter and Saturn are
expected to be present in other planetary systems and to be ubiquitous
as the gaseous envelopes of satellites and grains in the early epochs
of planet formation.  These extended atmospheres should be more
readily detected than the moons or rings of a giant extrasolar
planet. Such a detection would, therefore, provide a means for
determining the presence and the composition of satellites or rings
orbiting an extrasolar planet. The possibility of detecting gas from a
star-grazing comet orbiting an extrasolar body has been considered but
not yet observed \citep{men04}. However, sodium \citep{cha02},
hydrogen, oxygen, and carbon \citep{vid04} have been seen in
absorption in the atmosphere of the Jupiter-sized extrasolar planet
HD~209458b.

In this paper we first review the properties of the toroidal
atmospheres of Jupiter and Saturn, and use the observations of the
atmosphere of HD 209458b as a measure of the amount of absorbing
material that can be detected in a transit.  We then 
examine under what circumstances giant toroidal
atmospheres, produced by orbiting moons or ring particles, might be
observed around an extrasolar planet.
 
\section{Toroidal Atmospheres at Jupiter and Saturn}
 
Jupiter's closest moon Io has lost most of its early volatile
inventory, but it has a thin, volcanically produced atmosphere
composed primarily of SO$_2$ with other trace gases
\citep[e.g.,][]{mcg04}.  The atmosphere is stripped by interaction
with ions trapped in Jupiter's magnetosphere to produce an extended
toroidal atmosphere of neutral atoms and ions. Due to its strong
resonance fluorescence in the visible, sodium was the first species
detected in this extended atmosphere \citep{bro74}. The cloud of
escaping sodium shown in Fig.~1a co-orbits with Io and is now nearly
continuously imaged from earth. Using the stripping rate of sodium
$\sim 3\times10^{26}$~atoms~s$^{-1}$ and the average electron impact
lifetime $\sim$2~hr, this Jupiter-sized cloud contains $\sim 2 \times
10^{30}$ \ion{Na}{1} atoms \citep[e.g.,][]{tho04}. The sodium that is
ionized in this cloud is confined by the local magnetic fields and
contributes to the ambient plasma. It is eventually lost, primarily by
charge exchange, producing energetic neutrals that form a giant sodium
nebula with a radial extent of $\sim$500 Jupiter radii ($R_J$)
containing $\sim$10$^{32}$ \ion{Na}{1} atoms \citep{men90}.
 
The total rate of stripping from Io's atmosphere is more than 100
times that for sodium. The principal species ejected are sulfur and
oxygen, which have longer lifetimes than sodium. This material also
orbits Jupiter forming a giant torus of neutrals with a density and
spatial distribution determined primarily by electron impact
ionization.  The neutral sulfur and oxygen tori so formed have cross
sectional areas comparable to Jupiter's and contain $\sim 3 \times
10^{32}$ \ion{S}{1} atoms and $\gtrsim 1 \times 10^{33}$ \ion{O}{1}
atoms \citep{smy03}. Because Io is embedded deeply in Jupiter's
magnetic field, the ionized plasma formed from the orbiting neutrals
is also confined, leading to peak ion densities $\gtrsim
2000$~cm$^{-3}$ \citep[e.g.,][]{tho04}.  This toroidal plasma has been
imaged in emission in lines of \ion{S}{2}, \ion{S}{3}, and
\ion{O}{2}. The spatial distribution of the \ion{S}{2} 6731~\AA\
emission is shown in Fig.~1c.  Since the plasma lifetimes are much
longer than the neutral lifetimes, the plasma torus has a
significantly larger total content than the neutral torus $\gtrsim 3
\times 10^{34}$ ions, with an average column density $\sim 2 \times
10^{14}$~cm$^{-2}$ and a cross sectional area comparable to
Jupiter's. The torus of ejected material from the surface of Jupiter's
moon Europa has a comparable content of water products from Europa's
icy surface \citep{bur04,han05}.
 
The more distant giant planet Saturn has a small icy moon Enceladus
that is ejecting ice grains \citep{spa05} and gas, primarily H$_2$O
with trace carbon and nitrogen species \citep{wai06}. The orbiting
grains make up Saturn's tenuous E-ring that extends from $\sim$3 to 8
Saturn radii ($R_S$) \citep{sho91}. The gas ejected from Enceladus and
the sputtering of the E-ring grains create a large toroidal atmosphere
of H$_2$O \citep{joh06a,jur02} and its products, initially detected by 
the Hubble Space
Telescope \citep[Fig.~1b,][]{she92,jur02}.  This atmosphere also has a
toroidal ionized plasma, composed principally of H$_3$O$^+$,
H$_2$O$^+$, OH$^+$, O$^+$ and H$^+$ \citep{you05}. However, unlike the
Io torus, the average neutral density is larger than the ion density.
The neutral toroidal cloud contains $\sim 10^{35}$ OH molecules and
comparable amounts of H$_2$O, O, and H, with an average line of sight
column density $\sim 10^{15}$~cm$^{-2}$ over a region of radius $\sim
1$~$R_S$ (see Fig.~1b). Saturn's giant moon Titan also produces a
toroidal cloud \citep{smi04} of nitrogen, methane, and
hydrogen. Because it resides in Saturn's $\it{outer}$ magnetosphere,
the atmospheric stripping rate is low \citep{joh04a,mic05} and the ions
formed are rapidly lost down the magneto-tail. Therefore, due to its
location in the magnetosphere, Titan's toroidal cloud is not nearly as
robust as the water product torus in the inner magnetosphere. Although
the atmospheres described above were detected in emission, we consider
the likelihood of seeing such features in absorption associated with a
giant planet transiting the disk of its parent star. However, we first
consider the observations of the HD~209458b transit as an example of
the size of the absorbing column of gas that can be detected.

\section{HD 209458b Observations}
 
Sodium has been observed at the extrasolar planet HD~209458b in
absorption when the planet crosses the disk of its parent star
\citep[][]{cha02}. This sodium is likely to be a component of the planet's
gravitationally bound atmosphere (e.g., Charbonneau et al. 2002, 2006)
or its escaping atmosphere (Vidal-Madjar et al. 2004).  The sodium
absorption feature detected corresponds to a small change in intensity
($\Delta I/I = 0.0232$\%) at the 5890, 5896~\AA\ doublet
\citep{cha02}, and we use this to estimate the minimum number of
sodium atoms in a cloud that are needed to produce an absorption feature 
at this level, realizing that new techniques will soon be available 
(e.g., \citet{arr06}). Ignoring the effects of the
background stellar spectrum, the absorption signal can be expressed as
$\Delta I/I \sim W_{\lambda} / \Delta\lambda$, where $W_\lambda$ is
the equivalent width and $\Delta\lambda$ is the observing width $\sim
12$~\AA.  In the optically thin, linear absorption regime, the
equivalent width is given by the expression $W_\lambda = 8.9 \times
10^{-5} N_{ave}\lambda^2 f_{ik}$~\AA, where $N_{ave}$ is the average
column density of the absorbing species and $f_{ik}$ is the oscillator
strength: 0.641 and 0.320 for the 5890\,\AA\ and 5895\,\AA\ lines,
respectively \citep{san05}. The minimum, average column density,
N$_{ave}$, associated with this feature is therefore $\sim 1.2 \times
10^{10}$~cm$^{-2}$, spread over the area of the stellar disk ($\sim 2
\times 10^{22}$~cm$^2$). If the sodium is a component of the planet's
atmosphere, the absorbing area is very small, so that the
actual column density is large and the line is heavily saturated. If,
however, the sodium were in an extended cloud of radius one or a few
planet radii, the actual attenuating column density would be $\sim
10^{11}$--$10^{12}$~cm$^{-2}$. The total minimum number of sodium
atoms in the linear regime is $\sim 2 \times 10^{32}$. Sodium is,
however, a trace gas, and there are much larger amounts of more
abundant species.
 
\citet{vid04} observed \ion{H}{1}, \ion{O}{1}, and \ion{C}{2} at
HD~209458b by detecting a reduction in the parent star's emission
features in the UV during transit. They find that their results can be
understood if the absorbing material has a velocity dispersion
comparable to or greater than the widths of the stellar lines ($\sim
15$~km~s$^{-1}$ and 25~km~s$^{-1}$ for \ion{O}{1} and \ion{C}{2},
respectively) and conclude that the absorbing species are components
of an escaping atmosphere. For the observations of the 
\ion{O}{1} 1302 \AA\ line, Vidal-Madjar et al.\
show that the collisionally excited 1304~\AA\ and 1306~\AA\ lines also
contribute to the absorption. For simplicity we assume that the lines
contribute equally to the absorption and that the band depth of $\sim$0.13
applies to each one. A rough lower limit to the column of absorbing 
material can be made by writing the average absorption cross section as 
$\sigma \thickapprox a f_{ik} / \Delta\nu$ with a $\thickapprox$ 0.027 
cm$^2 s^{-1}$. Using 15~km~s$^{-1}$ for the stellar line widths,
$\Delta\nu = 1.1\times 10^{11}$~s$^{-1}$ and $f_{ik} \thickapprox 0.05$
\citep{san05}, so that $\sigma \thickapprox 1.2 \times 10^{-14}$ cm$^2$. 
The number of absorbing atoms in each sub-level is then $\sim 0.13\, 
A_*/\sigma$, where $A_*$ is the area of the stellardisk, which gives 
$\sim 2.3 \times 10^{35}$ \ion{O}{1} atoms. If they have a radial 
distribution of a few planet radii, the corresponding \ion{O}{1} column 
density is $\sim$10$^{14}$ cm$^{-2}$. In a low
density extended atmosphere, the sub levels above the ground state will not
be excited, so the single line values are appropriate. Similarly, for the
observations of the \ion{C}{2} 1335~\AA\ line, the excited 1336~\AA\ line
also contributes. Treating this in the same way, with a band depth of
$\sim$0.075 and $\it{f}_{ik}$ $\thickapprox$ 0.12 \citep{san05}, implies 
$\sigma \thickapprox2\times 10^{-14}$~cm$^2$ and a total content $\sim 
8\times 10^{34}$ \ion{C}{2} ions, which gives a similar column density.

\section{Toroidal Atmospheres around Extrasolar Planets}

We expect that extended atmospheres like those observed around Jupiter
and Saturn are present around gas giants in other planetary systems
and are likely to be ubiquitous as the gaseous envelopes of satellites
and grains in the early epochs of planet formation.  The toroidal gas
clouds around Jupiter and Saturn are seen in emission, excited by
resonance fluorescence or electron impact. Because of the extremely
low density, the atoms and ions are typically in the ground
state. In an extrasolar planetary system these extended atmospheres
could give rise to absorption spectra when the planet
transits the disk of the star.

In order to assess the feasibility of observing the extended
atmospheres of exoplanets we compare the content of these atmospheres
in Jupiter and Saturn with the minimum content needed to account for
the absorption features measured for HD~209458b as it transits its
parent star.  The minimum number of \ion{Na}{1} atoms that contribute
to the absorption feature detected at HD~209458b, $\sim 2 \times
10^{32}$, is approximately two orders of magnitude larger than that in
the sodium cloud immediately surrounding Io. It is, however,
fortuitously close to the number of \ion{Na}{1} atoms in the very
extended Jovian sodium nebula.  Similarly the minimum amount of
\ion{O}{1} at HD~209458b, $\sim 2 \times 10^{35}$ atoms, is about two
orders of magnitude larger than the \ion{O}{1} content in Io's torus,
but it is comparable to the amount of oxygen containing species in
Saturn's more substantial neutral torus. In addition, the
\emph{ionized} minimum carbon content at HD~209458b $\sim 8\times
10^{34}$ \ion{C}{2} ions, is comparable to the sulfur and oxygen 
$\emph{ion}$ content in the Io plasma torus. Therefore, detection of 
sulfur ions at this level during a planet transit could be evidence 
for the presence of an Io-like moon.

Thus the ion or neutral content of a toroidal atmosphere, produced by
a satellite or ring about a transiting extrasolar planet, could be
sufficient to give a detectable absorption feature.  The steady state
densities formed in the vicinity of a transiting giant planet are
dependent on a number of different factors, and we consider below
whether there are situations in which toroidal atmospheres like those
around Jupiter or Saturn might be seen around a giant planet
transiting its parent star.

\subsection{Orbit Size and Stability}

The possibility of observing a toroidal atmosphere in absorption when
a giant planet transits the disk of its parent star is determined not
only by the amount of material and the dimensions of the cloud, but
also by the geometry of the observation. The orbital period of planet
HD~209458b is short, 3.5 days. Its orbital plane cuts across our line
of sight to the parent star, and repeated transits have been observed
\citep[e.g.,][]{cha00,wit05}. A planet with an orbital plane that is
randomly oriented with respect to the line of sight to the star will
have a probability $\sim R_*/a$ of transiting the disk of its star,
where R$_*$ is the radius of the star and $a$ is the radius of the
planet's orbit, which we assume to be circular \citep[]{bor84}. Small
orbital radii are therefore favored for observation. The probability
that a planet like HD~209458b would be seen transiting the disk of its
parent star is $\sim$0.1.

When the orbital radius of a planet is small it can become phase-locked to
its star and its ability to maintain satellites in stable orbits is reduced
\citep{bar02}. HD 209458b, for instance, is thought to be phase-locked to
its star [e.g.,\citet{sea02,gri04}] and its Hill sphere is only about 4.3
times its radius. Barnes and O'Brien (2002) estimate that an Io-like moon
would last only $\sim 2\times 10^6$~yr around HD 209458b and the largest
satellite that could survive would have a radius $\sim 1/40$ of that of Io.
Therefore, although giant planets certainly had orbiting material early in
their history, they are not likely to maintain large satellites if they
orbit close to their parent star. Using conventional ideas on tidal
locking, a Jupiter-sized planet orbiting at $\sim$0.05~AU could become
phase locked to a solar type star in $\sim 2\times 10^6$ yr
\citep{gui96,sea02}. This time increases to $\sim 1 \times 10^8$ yr at
0.1 AU and $\sim 7\times 10^9$ yr at 0.2 AU. The planet's Hill sphere
also increases with the distance from the star.  Therefore, although
toroidal atmospheres associated with orbiting debris might be detected
early in the history of a giant planet at a small orbital radius, we
focus on the possibility of detecting toroidal atmospheres associated
with planets having orbital radii $\gtrsim 0.1$~AU.

\subsection{Structure of a Toroidal Atmosphere}

The total content, morphology, and kinematics all affect the
absorption characteristics of an extended atmosphere. These depend on
several parameters including the source rate ($S$) and lifetime
($\tau$) of the species of interest, as well as the ejection velocity
($v_e$), and the orbital velocity of the source ($v_o$).  The total
number of atoms or molecules in a neutral cloud depends on the product
$S\tau$. Atoms and molecules ejected from an orbiting moon in a region
where $\tau$ is much shorter than the orbital period, $\tau_0$, will
form a neutral cloud with a morphology resembling the sodium cloud at
Io shown in Fig.~1a. The mean radial extent of the cloud is given by
$R_c \sim v_e\tau$ so that the cloud has a cross section of $A_c \sim
\pi R_c^2$, and the average column density is $N \sim S / \pi v_e^2
\tau$. A cloud with a toroidal morphology is formed if the atoms and
molecules are ejected from a continuous ring of material \emph{or}
they are ejected from an orbiting moon \emph{but} with lifetimes,
$\tau \gg \tau_o$. A toroidal atmosphere of ionized gas can be formed
from the ions produced by \emph{either} short \emph{or} long-lived
neutrals that are picked-up and distributed azimuthally by the
rotating magnetic field, which is the case for the \ion{S}{2} torus
shown in Fig.~1c.
 
In order to illustrate the characteristics of a neutral torus, we show
in Fig.~2 the line of sight column densities calculated using a Monte
Carlo model \citep[e.g.,][]{jur02,joh06a} for an outgassing moon orbiting a
planet with the mass of Saturn. The orbit radius $R_{ps} = 3.9$~$R_S$,
the orbital velocity $v_o= 12.6$~km~s$^{-1}$, the escape velocity $v_e
= 2$~km~s$^{-1}$, and the product $S\tau = 10^{35}$ atoms. These
quantities correspond to the orbital parameters of the moon Enceladus
and the content of its neutral toroidal atmosphere.  The column
densities in Fig.~2 are obtained by viewing the torus edge on. This is
approximately the alignment expected in a transit because the planet's
rotational axis is typically aligned with its orbital axis and most known 
satellites and all rings orbit near their planets' equatorial planes. 
Small tilts in the orbital plane will reduce the line-of-sight column
density somewhat. The neutrals in this torus have an average speed
roughly equal to the orbital speed, which for Enceladus is $\sim
13$~km~s$^{-1}$. The plasma formed from ionized neutrals would
co-rotate with the planet, which corresponds at the orbital radius of
Enceladus to a speed of $\sim 40$~km~s$^{-1}$. These speeds would
result in line widths that are comparable to the stellar line widths
discussed in \S3 for the HD~209458b observations.  If the absorption
lines could be observed at high spectral resolution, the change in line
profile at the beginning and end of the transit would be a valuable
diagnostic.
 
The neutral content of the extended atmosphere is given by the
quantity $S\tau$, as noted above, but the vertical and radial extents
of the torus depend on the ratio of the mean ejection speed, v$_e$,
and the orbital speed, v$_o$.  When $v_e/v_o$ is small, which is the
case for the model in Fig.~2 and for the features in Figs.~1a--c,
centrifugal confinement dominates. That is, in a collision-less torus
the vertical and radial extents of the cloud are determined by the
distribution in the eccentricities and inclinations of the orbiting
neutrals. For small $v_e/v_o$, the maximum excursion above or below
the orbital plane is approximately equal to the vertical component of
v$_e$ divided by the angular velocity of the source, $v_o/R_{ps}$,
where $R_{ps}$ is the distance from the planet to the ring or
satellite source. Using isotropic emission from the source, the
average vertical distribution about the orbital plane is roughly $H
\sim R_{ps}v_e/v_o$. Similarly, the radial extent along the orbit
plane is $\sim 4H$ \citep[e.g.,][]{joh90,joh06a}. Observing a torus 
with an average circumference $2 \pi R_ {ps}$ seen edge on, the extent 
along the line of sight through the source region is $\sim
4(R_{ps}H)^{1/2}$. Therefore, an average column density $N \sim S \tau
/ 2 \pi R_{ps}^2 (v_e/v_o)^{3/2}$ can be observed over an area $A \sim
4H^2 = 4 R_{ps}^2(v_e/v_o)^2$ centered on the source. Using the
parameters of the model given above, this expression gives a column
density $\sim 10^{14}$~cm$^{-2}$ roughly consistent with the
simulation shown in Fig. 2. For this toroidal atmosphere, the average
column density and area over which the gas is distributed are
comparable to that for the atmosphere detected at HD 209458b.

In the simulation shown in Fig.~2, the initial molecular motion is
controlled by $v_e$, the mean ejection speed. If, however, the
neutrals are long-lived, they can be heated by interaction with the
co-orbiting plasma, which expands the cloud, as is the case for the OH
torus shown in Fig.~1b. We also note that $v_o$ varies as
$R_{ps}^{-1/2}$, so that toroidal neutral clouds formed at smaller
orbital radii give higher line-of-sight column densities, and hence, 
deeper absorption features.
 
\subsection{Scaling to Smaller Orbits}

With the equations given above, the content, column density, and size
of a neutral torus can be roughly scaled for planets with  
different orbits using
estimates of the source rate, mean lifetime, mean ejection speed,
orbital location, and stellar luminosity. Motivated by the
considerations given in \S4.1, we use this scaling to consider if an
extended atmosphere around a giant planet closer to its parent star
than Jupiter or Saturn could be observed in a transit.  As an example,
we consider a planet with an orbit of radius $\sim$ 0.2~AU that might
have a torus with absorption features comparable in strength to those 
observed in the atmosphere of 
HD~209458b. At a distance of 0.2~AU, the probability that the planet
would transit the stellar disk along our line of sight is $\sim
0.02$.  The detection of transits at such orbital distances requires
searches of significant numbers of stars over an extended time because
of the relatively long orbital periods.(e.g., Sozzetti et al. 2004;
Pepper \& Gaudi 2005).

As described earlier, the origin of Io's atmosphere is volcanism,
produced by tidal heating of the interior. The supply of gas to the
atmosphere depends on Io's position relative to its parent planet and
its companion moons, but not on the distance from the Sun. However,
the size and content of the Io's atmosphere, which affects the
stripping rate, does depend on the orbital radius of the parent planet. 
If the Jovian system were moved closer to the Sun, to an orbit with 
radius $\sim 0.2$~AU, the increased temperature would increase the 
atmospheric loss rate to a level approaching the volcanic gas production 
rate, about an order of magnitude larger than the present source rate 
for the Io torus. However, the increased radiation field at 0.2~AU 
would decrease the \ion{Na}{1} photo-ionization lifetime by a factor 
$\sim 600$.  The net effect is that the neutral sodium cloud would 
become more tenuous than the current cloud around Io.  However, this 
conclusion needs to be modified if the star has a spectral type much 
later than G2 so that the \ion{Na}{1} photo-ionization rate is 
significantly lower than in the solar system (the \ion{Na}{1} absorption 
edge lies at 2400~\AA), or if the object is observed at an early epoch 
during the loss of the moon's primordial atmosphere when the source 
rate is expected to be significantly larger \citep[e.g.,][] {joh04a}.

Oxygen in the toroidal cloud at Io is ionized by electron impact with
a lifetime $\sim$ 10$^5$~s. When the neutral source rate is very
large, however, the electrons are rapidly cooled and photo-ionization
becomes the limiting factor.  Photo-ionization is also the dominant
destruction mechanism if the planet is close to the star.  The average
photo-ionization lifetime of oxygen at 0.2~AU is $\sim 10^5/r_L$~s
\citep{hue92}, where $r_L$ is the ratio of the photo-ionizing
luminosity of the star to the photo-ionizing luminosity of the Sun.

For a solar-type star, this is comparable to the present electron
impact lifetime at Io.  For an \ion{O}{1} torus at $\sim 0.2$~AU, a
cloud of the $\sim 10^{35}$ atoms extending $\sim$1~$R_J$  would
require a source rate of $S \sim 10^{30}r_L$~atoms~s$^{-1}$. This 
would require an escape rate $\sim 5 \times
10^{11} r_L$~atoms~cm$^{-2}$~s$^{-1}$. For $r_L = 1$ this is only
about 10 times the current atomic loss rate at Io which has, at
present, a thin atmosphere. Therefore, such a loss rate is certainly
feasible in an early epoch for a moon like Io with a thick Titan-like 
atmosphere and an exobase altitude that is a significant fraction of 
its radius. At this rate, a thick, primordial atmosphere (with $7\times
10^{27}$~amu~cm$^{-2}$) would be lost in about $3\times10^7
r_L$~yr. This is consistent with loss rates for the early atmospheres
of the Galilean satellites of Jupiter \citep{joh04a}, but would require
observing the planetary system at an early epoch.  Due to its much
larger surface area, the erosion of a ring of debris could more
readily supply such a toroidal atmosphere, as could volcanic emission
or out-gassing from a small body for which escape is not significantly 
limited by gravity, such as an Enceladus-like object.

The magnetic field is crucial for the maintenance of a toroidal
ionized plasma, but the presence of a magnetic dynamo on a planet with
a small orbital radius is uncertain \citep{gri04}. For the orbital
radii that we are considering ($\gtrsim 0.1$~AU), planets are not
likely to be phase-locked to their star so their magnetic field
strength could be comparable to that of Jupiter. In this case the
ability to confine the ions would also be similar to that at Io.
Based on the Jovian torus observations discussed in \S2, ion lifetimes
are determined by plasma diffusion processes and, therefore, are less
affected by the photo-ionization flux at a planet with a small orbital
radius. Using the average ion lifetimes near Io, the required source
rate would be $\sim 3\times 10^{28}$~atoms~s$^{-1}$, much smaller than
the case discussed above for producing an observable neutral
torus. This rate is comparable to the present out-gassing rate of
Enceladus and to the atmospheric stripping rate at Io. Therefore, if
the planet has a robust magnetic field, an Io-like moon can produce a
toroidal \emph{plasma}, like that in Fig.1c, which has a content and
scale roughly comparable to the plasma feature already observed in the
atmosphere of HD~209458b.
  
A toroidal atmosphere like that at Saturn, formed from the out-gassing
of an icy moon, is unlikely to be present around a solar-type star at
an orbital radius much less than $\sim$5~AU because the ice mantel
could not survive. The probability of observing a transit in this case
is therefore low. Recently, however, a planet candidate has been seen
orbiting the young brown dwarf, 2M1207 (Chauvin et al., 2004),
suggesting the presence of planets around low luminosity stars. By
scaling from the solar system, we find that a toroidal atmosphere
produced by out-gassing from an Enceladus-like object could be present
on a planet with an orbit of radius 0.2~AU around a star of much lower 
luminosity
$\sim 1.5 \times 10^{-3}$~$L_\odot$.  The measurement of the spectral
signature of the torus becomes correspondingly more difficult in the
low luminosity system, but intermediate cases could be observable,
especially where the source rate is significantly larger than for
Enceladus, which is easily feasible especially at early epochs.

\section{Summary}
                                                                               
The giant planets Jupiter and Saturn have large emission features
associated with ejecta from satellites and ring particles. These
ejecta can form clouds of neutrals and ions that have cross sections
comparable to the planet, as seen in Fig.~1. We point out that such
features in our solar system have sizes and contents comparable to the
escaping atmosphere already observed in absorption when HD~209458b
transits its parent star. Therefore, we suggest that, although it is
very difficult to observe a moon or ring on an extra solar planet,
it might be more likely to observe the torus of gas escaping from a
satellite or a ring of debris when a giant planet makes a transit
across the disk of its star.

Since the probability of observing a transiting planet increases with
decreasing distance from the parent star, detection of giant planets
with small orbital radii are favored. Such planets very likely had 
toroidal atmospheres from orbiting debris or satellite venting early
in their history. For instance, the large Jovian satellites, which
have been fully stripped of their primordial atmospheres
\citep[e.g.,][]{joh04a}, likely had robust atmospheres at earlier
epochs. However, for a planet with a small orbital radius, like HD
209458b, maintaining satellites in stable orbits is problematic
\citep{bar02}.

For stars with giant planets that have larger orbital radii, an
observable toroidal atmosphere formed from refractory materials could
be present, as is the case at Jupiter. Also, a planet orbiting a much
cooler star at a relatively small radius could maintain a toroidal
atmosphere formed from icy bodies like that seen at Saturn. Detecting
such features on an extrasolar planet would be important. It would
indicate the presence of satellites or rings and, possibly, the
presence of a magnetic field.

\acknowledgments 
We thank J.-M. Grie{\ss}meier, D. P. O'Brien and the referee for many 
helpful comments, and Ms.~M. Liu for assistance in preparing the 
manuscript and figures. REJ also thanks the Physics Department of NYU 
for hospitality while on sabbatical leave. We acknowledge support from
NASA Planetary Atmospheres and NASA Origins programs (RJE), and NSF
AST 03-07277 (PJH).

\clearpage

\begin{figure}
\epsscale{.50}
\plotone{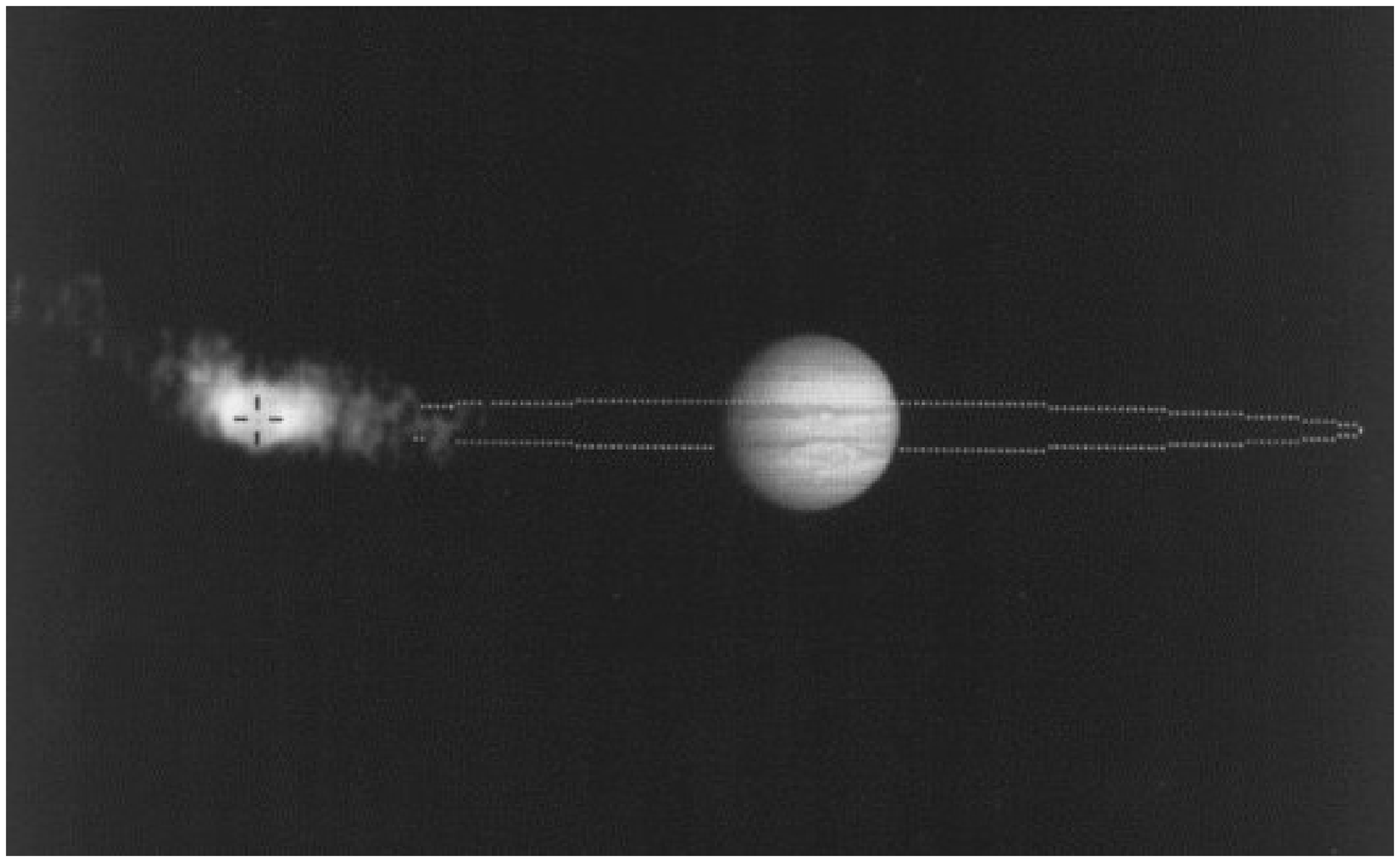}
\plotone{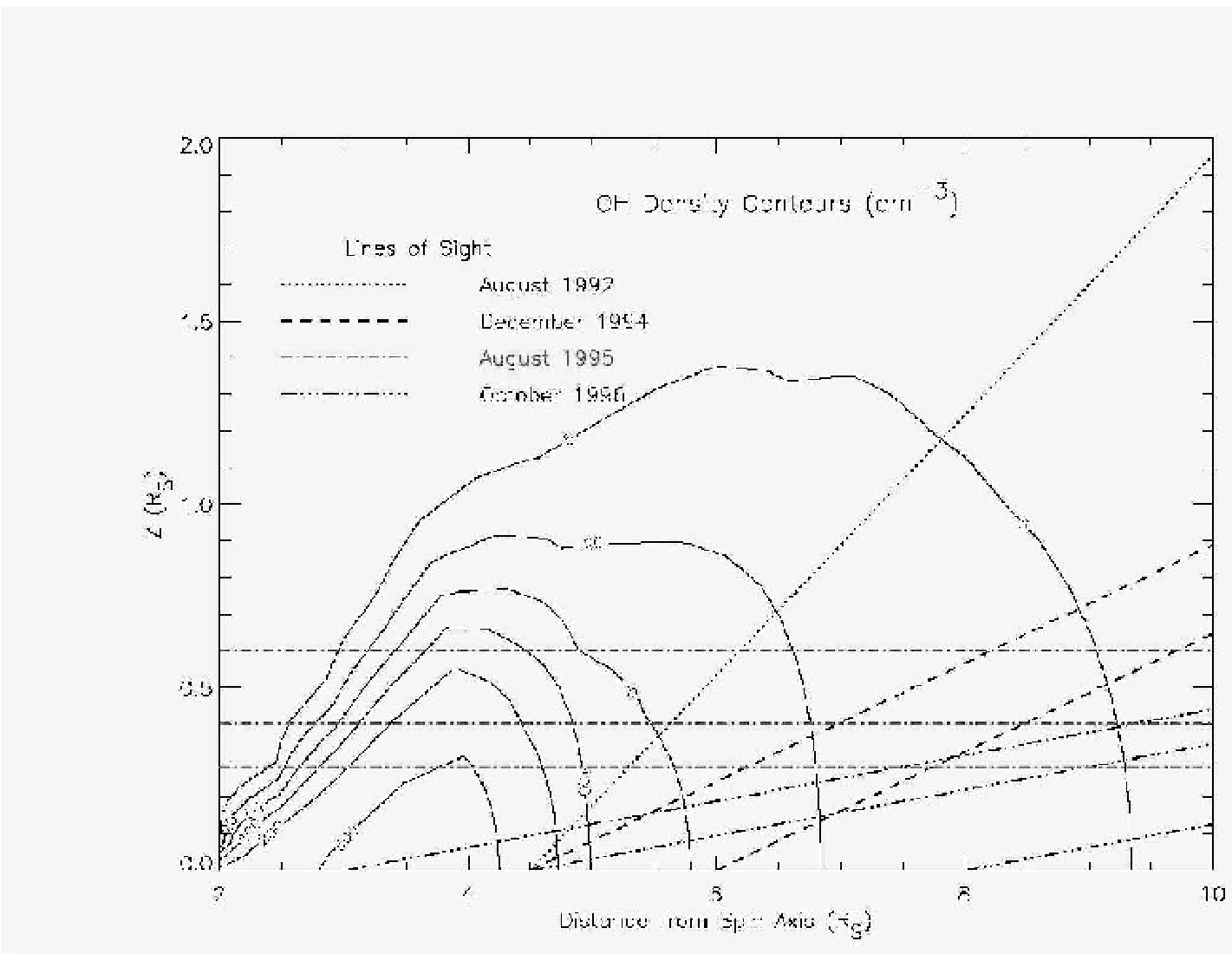}
\plotone{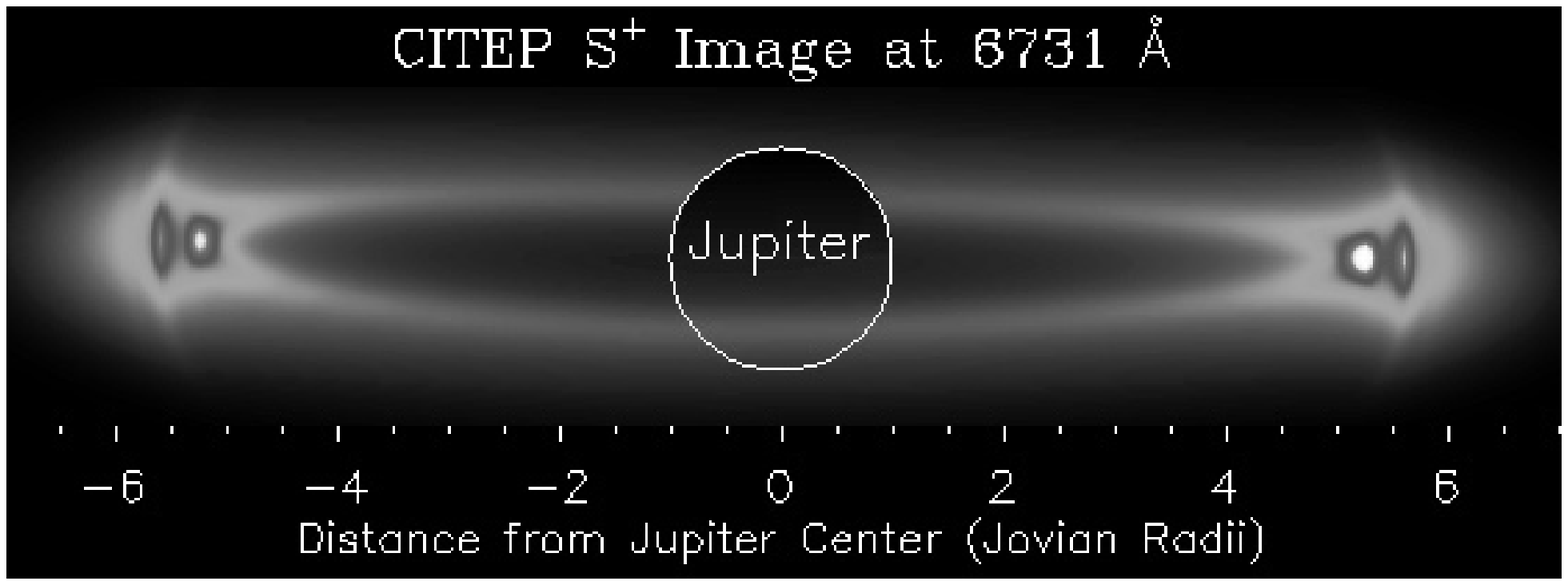} 
\caption{(1a) Image of the sodium cloud at Io in the D2 5890~\AA\
resonance florescence line. The image was taken with the 61-cm
telescope on Table Mountain \citep{gol84} and adapted from Fig.~4 of
\citet{smy92}.  The intensity is in Rayleighs ($4 \pi \times
10^{-6}$~photons~s$^{-1}$~cm$^{-2}$~sr$^{-1}$) with the peak at $\sim
7\times 10^3$~Rayleighs. The sodium is ejected from Io's volcanic
atmosphere by incident magnetospheric plasma ions and co-orbits with
Io until ionized.  The image of Jupiter is superposed, and the
continuous line indicates Io's orbit at 5.9~$R_J$ (1 $R{_J}$ =71,472
km).  
(1b) Model of the OH toroidal atmosphere at Saturn constructed
using the line of sight column densities measured using the Hubble
Space Telescope \citep{jur02}. The observational lines of sight are
indicated in the figure. The peak densities ($\sim$ 1000 cm$^{-3}$)
occur near the orbit of the moon Enceladus at $\sim 3.9$~$R_S$ (1
$R_{S}$ = 60,300 km). The neutral torus is formed from gas and ice
grains ejected by the moon.  
(1c) Image of the Io plasma torus in
emission in the \ion{S}{2} 6731~\AA\ line \citep{sch95}. The sulfur
(like sodium in Fig.~1a) originates from Io's volcanic atmosphere and
is ionized and picked-up by Jupiter's rotating magnetic field. The
emission is produced by electron impact excitation.}
\end{figure}

\clearpage
 
\begin{figure}
\epsscale{1.0}
\plotone{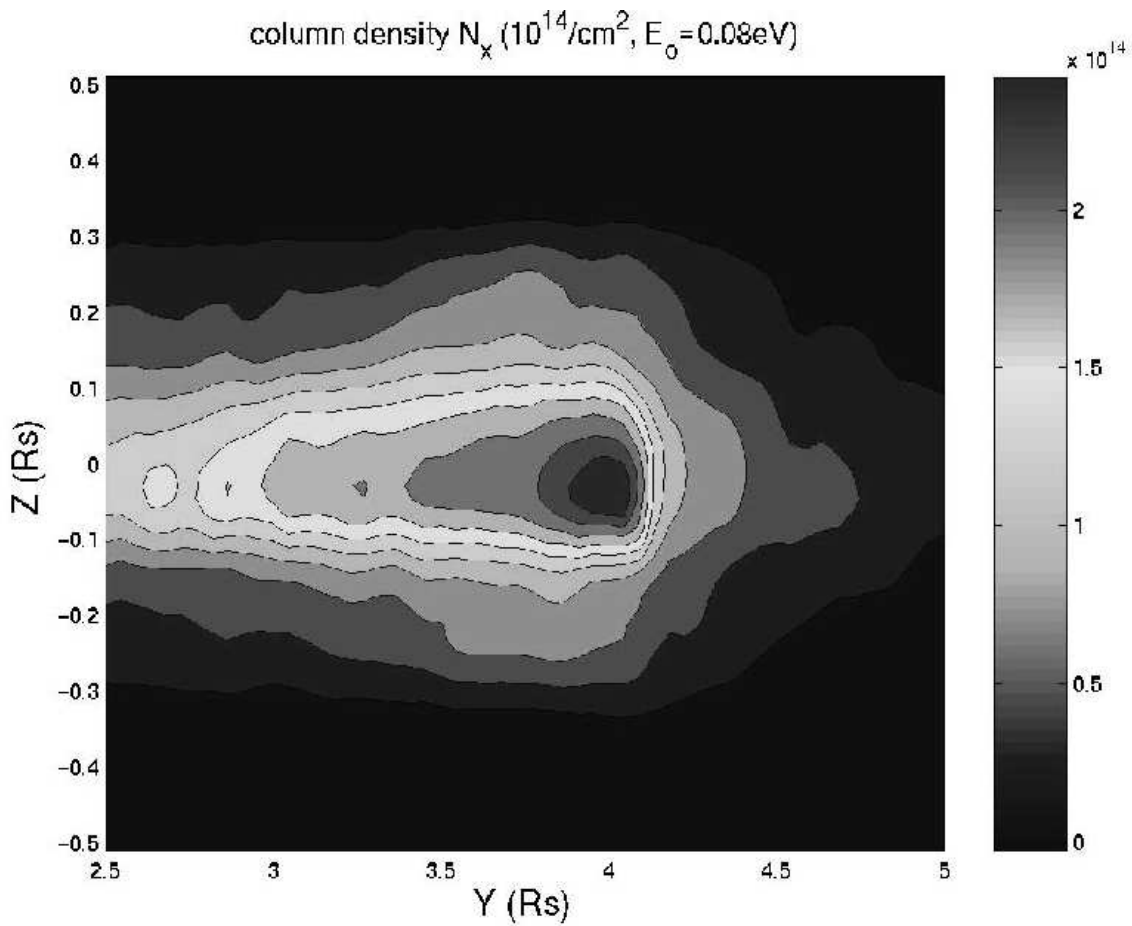}
\caption{ Monte Carlo model for the column density of gas in a
neutral torus formed from an out-gassing moon. The torus is seen
edge-on. The horizontal and vertical axes are the distance from the
planet in the orbital plane, and perpendicular to the plane,
respectively, in Saturn radii.  The orbital radius for the moon is
that of Enceladus, 3.9~$R_S$. The model parameters are: $S\tau =
10^{35}$, $v_e = 2$~km~s$^{-1}$, and $v_o = 12.6$~km~s$^{-1}$.  The
ratio $v_e/v_o$ determines the vertical and radial extents of the
torus.  See text for details.}
\end{figure}

\end{document}